\documentclass[12pt,twoside]{article}
\usepackage{fleqn,espcrc1}

\usepackage{graphicx}
\usepackage{psfig}
\usepackage[figuresright]{rotating}

\def\NJ{N_{\J}}
\def\J{J/\psi}

\newcommand{\AmS}{{\protect\the\textfont2
  A\kern-.1667em\lower.5ex\hbox{M}\kern-.125emS}}

\title{J/Psi Production at RHIC in a QGP\thanks{Supported in part by
U.S. Department of Energy Grant DE-FG02-95ER40213}
}

\author{Robert L. Thews\address[UofA]{Department of Physics, University of
Arizona, Tucson, AZ 85721, USA} and Johann Rafelski\addressmark[UofA]}
\begin{document}

\maketitle

\begin{abstract}
{In central collisions at RHIC, 
the initial production of heavy quarks will for the
first time yield multiple pairs of $c\bar{c}$
in each central event.
If a region of deconfined quarks and gluons is subsequently
formed, a new mechanism for the formation of heavy quarkonium
bound states will be activated.  This will result from the
mobility
of heavy quarks in the deconfined region, such that
bound states can be formed from a quark and an antiquark which
were originally produced in separate incoherent interactions.
Our model estimates of this effect predict a dramatic increase
in the number of observed $\J$ at RHIC, over that predicted from
extrapolation of color-screening or gluon dissociation mechanisms from
the lower CERN-SPS energies.  The centrality and energy dependence
of this effect should be readily observable by the Star and Phenix
detectors.  Thus the $\J$ abundance at RHIC will 
continue to provide a signature
of QGP formation. However, it is in this environment a more useful
probe, since contrary to prior expectations this large predicted
$\J$ abundance should be relatively easy to measure.

}
\end{abstract}

\section{Introduction}

The pattern of ``anomalous suppression'' of $\J$ 
observed by the NA50 experiment 
in Pb-Pb collisions at CERN \cite{NA50}
has been interpreted in terms of the effects of quark-gluon plasma
formation.\cite{Satz}  A straightforward extension of this scenario
to RHIC collisions would predict that virtually all of the 
initially-produced $\J$ would be suppressed, even for relatively
peripheral collisions.\cite{vogt,dinesh} In these pictures, the 
formation of $\J$ from a pair of charm quarks 
only occurs at the initial times.  Subsequent to that, 
the final state effects of interactions with nucleonic matter,
plasma screening, and dissociation by collisions with deconfined gluons
lead to the observed suppression.  
Recombination of the charm quarks into a $\J$ at hadronization is 
much less likely than formation of open charm particles, thus preserving
the suppression effect.  A tacit assumption in this chain of arguments
is that there is no recombination of charm quarks while in the deconfined
state.  While this may be approximately true in terms of numerical
significance, it is of course theoretically inconsistent to neglect the
recombination process.  For example, formation by capture from an
octet state with single gluon emission is precisely the inverse of
the primary gluon dissociation process.  The existence of both of these
processes depend on the ability of deeply-bound states such as the
$\J$ to exist above the deconfinement transition.\cite{karshsatz}

These considerations have the potential to become numerically 
significant at RHIC energies and above.  Estimates of initial
production from pQCD calculations indicate that approximately 
10 charm quark pairs will be present in each central Au-Au
collision at 200A GeV energy.  Since the formation rate will 
be proportional to the square of the number of unbound charm
quarks, there is a possibility that 
the recombination effect 
can increase dramatically.  It has been noted recently
that a statistical recombination model utilizing increased
heavy quark production at high energy will also result in 
formation of hidden-flavor enhanced by a quadratic dependence
on total heavy flavor.\cite{pbms,frankfurt} We emphasize that our
mechanism is quite distinct, in that a space-time
region of color deconfinement is a necessary condition for the
existence of our formation and dissociation processes.  

\section{Kinetic model}

We estimate the expected $\J$ population at RHIC in a detailed
kinetic model, in which the competing rates of formation and dissociation
are controlled by input densities.  

\begin{equation}\label{eqkin}
\frac{d\NJ}{d\tau}=
  \lambda_{\mathrm{F}} N_c\, \rho_{\bar c } -
    \lambda_{\mathrm{D}} \NJ\, \rho_g\,,
\end{equation}
where $\tau$ is the proper time,
 $\rho$ denotes
number density,
and the reactivity $\lambda$ is
the reaction rate $\langle \sigma v_{\mathrm{rel}} \rangle$
averaged over the momentum distribution of the initial
participants, i.e. $c$ and $\bar c$ for $\lambda_F$ and
$\J$ and $g$ for $\lambda_D$.

The gluon density is that of
an ideal gas in thermal and chemical equilibrium.  The initial 
charm quark population is due to initial production of charm quark pairs
according to pQCD calculations.  We assume one-dimensional isentropic
expansion to get a generic time-temperature profile.  
The formation process is radiative (gluon) capture from a c-cbar
pair in a color octet state.  
The formation process is just the inverse reaction, and
the final $\J$ population will be determined by the time integral of
the rates for these competing reactions.
(Note that in
this scenario the D mesons and other hidden charm bound states do
not appear in the kinetic equation, since their relatively small
binding energies prohibit their existence in a deconfined medium.)
For further details of our kinetic
model see~\cite{rlt}.
The magnitude of the formation term 
is quite sensitive to the momentum distribution of the charm quarks.
We consider a wide range of possible distributions, ranging from 
a thermal distribution at the plasma temperature to the initial
momentum distribution of the pQCD production process.

\section{Results}

\begin{figure}[htb]
\begin{minipage}[t]{80mm}
\psfig{width=7.5cm,figure=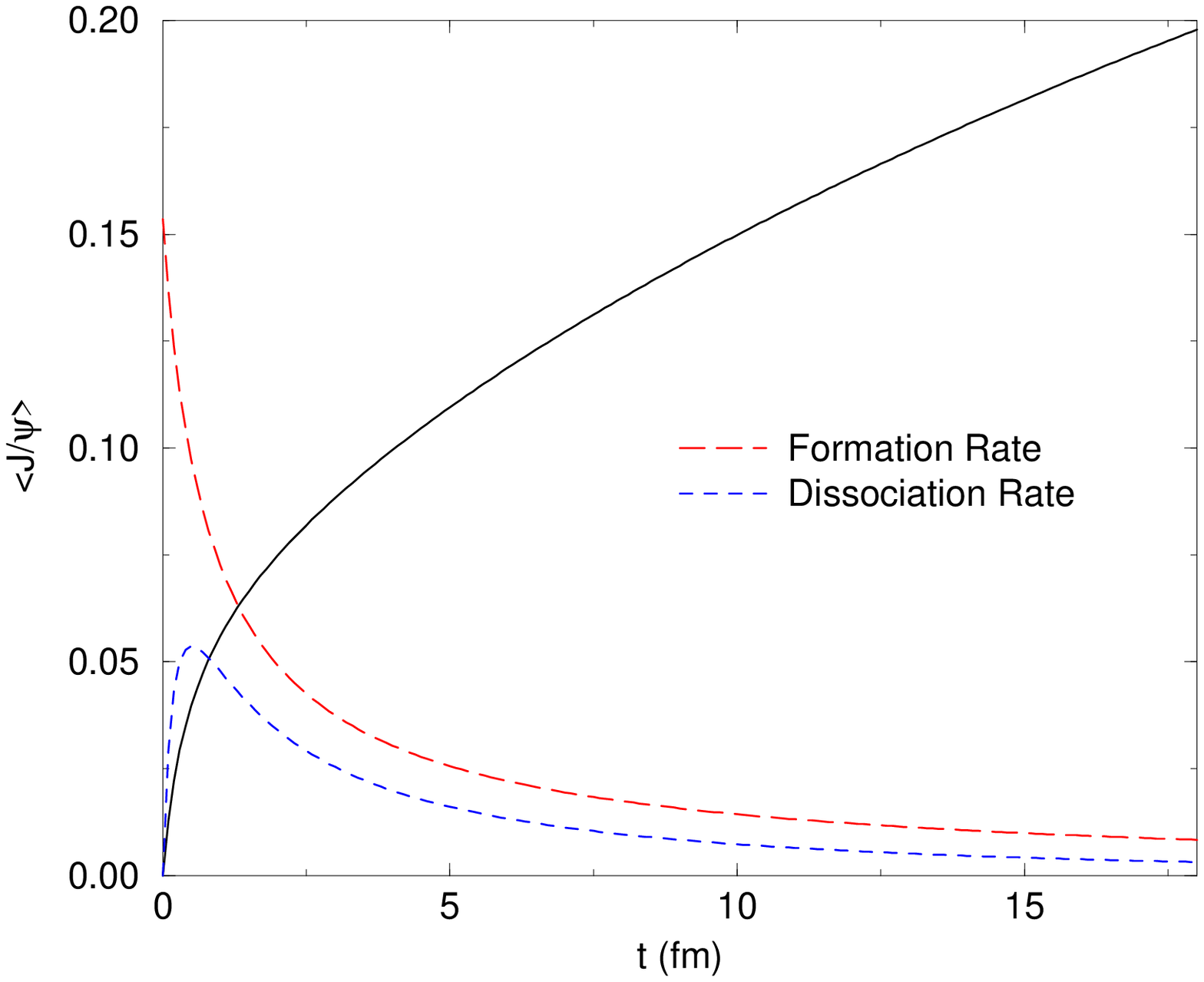}
\caption{\small Time development of the $\J$ population, and rates
of formation and dissociation.}
\label{jpsitimeplot}
\end{minipage}
\hspace{\fill}
\begin{minipage}[t]{75mm}
\psfig{width=7.5cm,figure=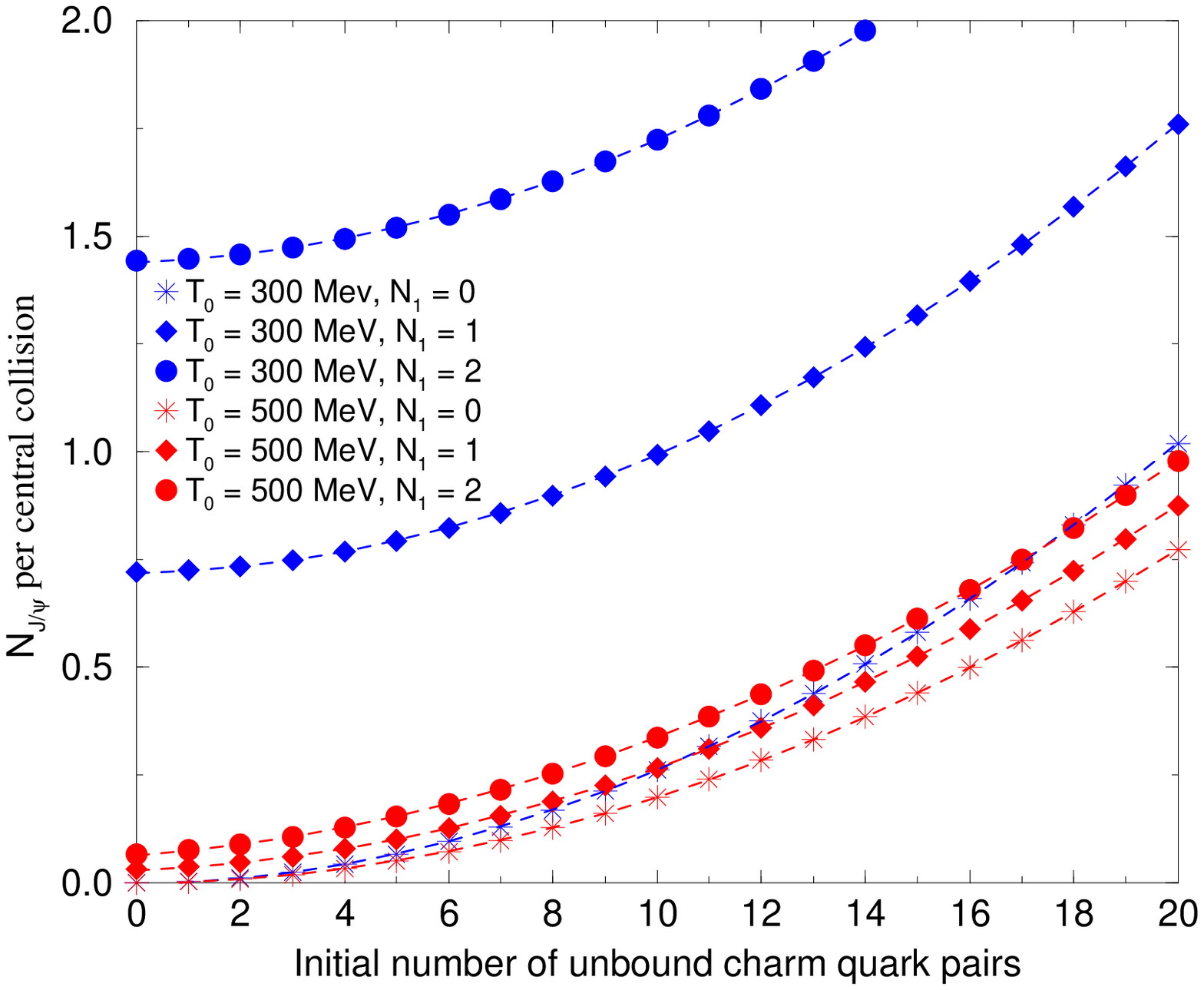}
\caption{\small Quadratic dependence of the
final $\J$ abundance on unbound charm.}
\label{quadraticfit}
\end{minipage}
\end{figure}

The time evolution of the $\J$ population is shown in Figure~\ref
{jpsitimeplot} for a typical choice of parameters.  Also shown are
the formation and dissociation rates.  One sees that the formation rate
drops due to the reduced charm quark density as the system expands, but
remains larger than the dissociation rate at all times.  This feature 
originates in the detailed balance factors which favor the exothermic
formation process.  Note that the final population of about 0.2 $\J$ 
per collision is greater than that expected from a simple superposition
of nucleon-nucleon interactions, typically about 1 \% of the 10 initial
charm quark pairs produced.   In Figure~\ref{quadraticfit} we 
verify the
anticipated quadratic dependence on the initial number of charm quark
pairs (exact charm conservation was enforced throughout the numerical
solution of the kinetic equations).   

\begin{figure}[h]
\begin{minipage}[h]{80mm}
\psfig{width=8.5cm,angle=0,figure=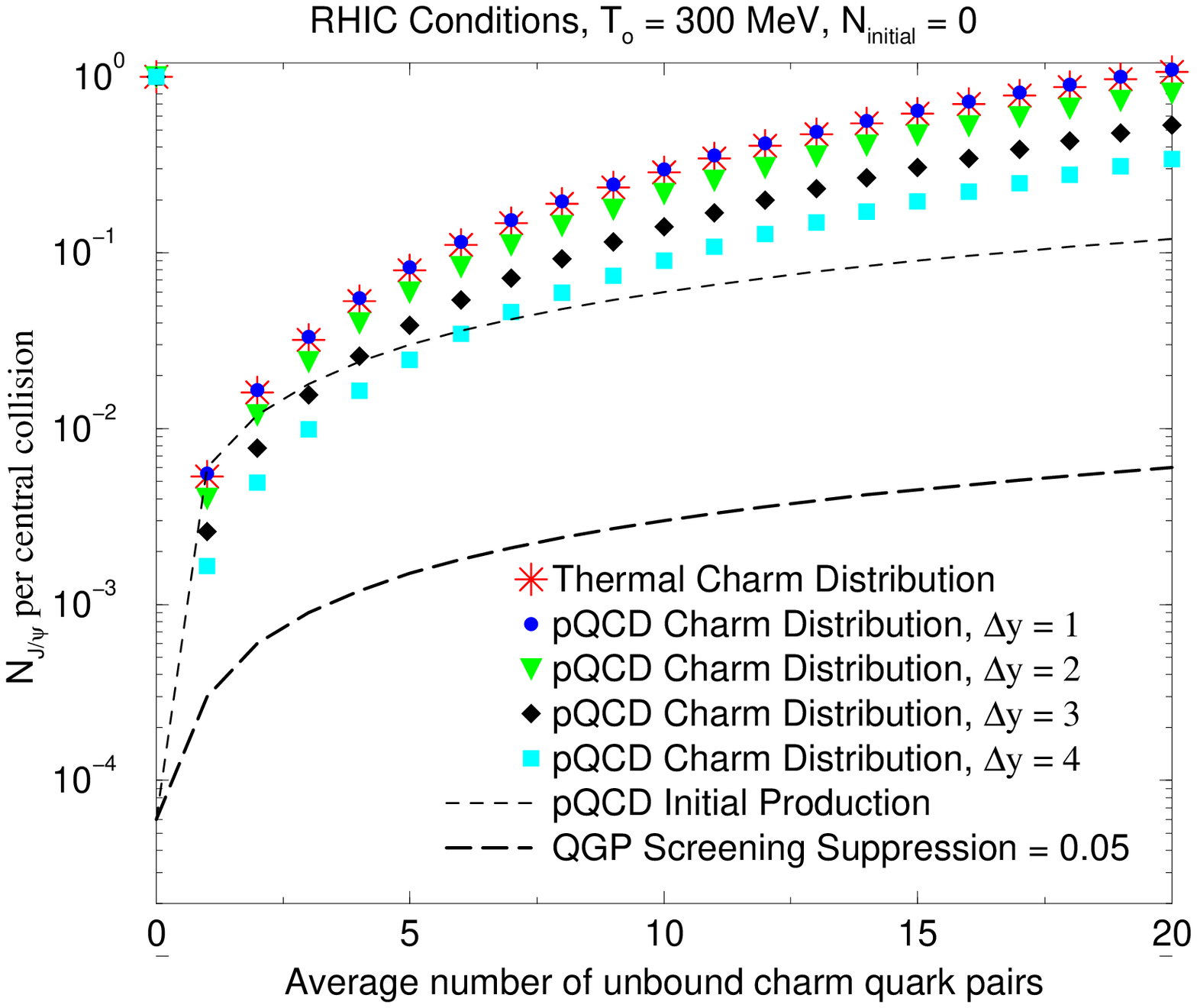}
\caption{ \small
Calculated $\J$ formation in deconfined matter for a range
of initial charm number and various charm momentum distributions,
for central collisions at RHIC at 200A GeV.
\label{jpsicharmdist}}
\end{minipage}
\hspace{\fill}
\begin{minipage}[h]{75mm}
\psfig{width=8.5cm,angle=-90,figure=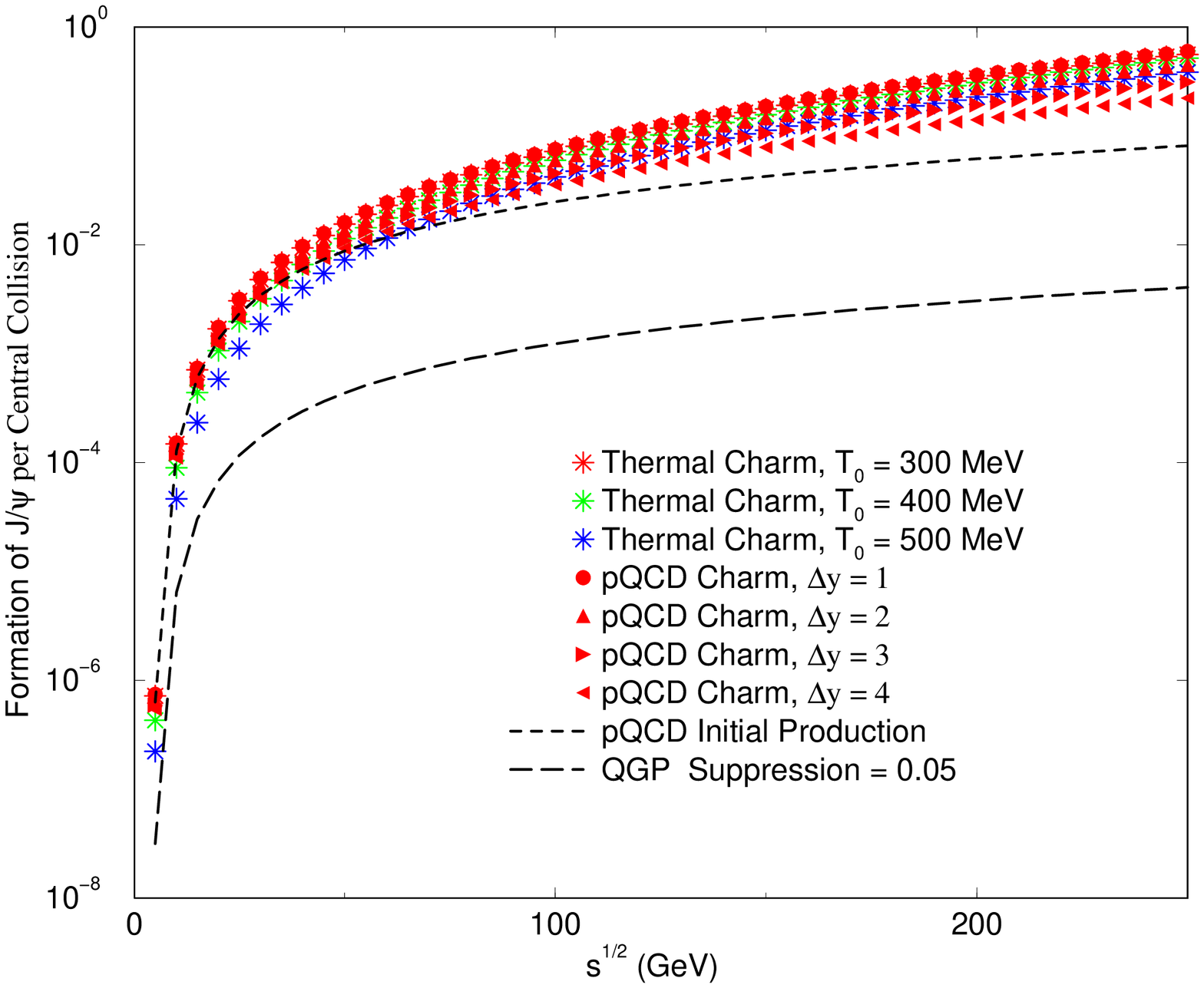}
\caption{ \small
Calculated $\J$ formation in deconfined matter for central
collisions at RHIC as a function of collision energy. No energy
variation of the plasma parameters is included.
\label{jpsienergy}}
\end{minipage}
\end{figure}

We then average over a distribution of initial charm and $\J$ to
obtain the final $\J$ population as a function of average initial
charm.  The results are shown in Figure~\ref{jpsicharmdist} for
various charm quark momentum distributions.  We parameterize those
from the initial pQCD calculations in terms of an effective rapidity
width $\Delta$y.  

One sees that even within the variation due to 
the momentum distribution, the final $\J$ population exceeds that 
expected from a superposition of nucleon-nucleon collisions (the
canonical 1 \% of initial charm indicated by the short dashed line), 
and is very far above that which would follow from a screening 
suppression factor typically 0.05 for RHIC (the long dashed line).

Variation with centrality can also be calculated in this model.  We
predict a unique behavior in this case, since the quadratic dependence 
on total charm produces an effect which grows with centrality.  This 
is in strong contrast with a pure screening scenario, in which 
the maximum suppression occurs for central events.  For details, see
Ref.~\cite{rlt}.

Finally, we consider the energy dependence within the RHIC range.
The effect here is due to the energy dependence of the initial
quark production, which is not far from linear in this limited range.
The results are shown in Figure~\ref{jpsienergy} for a full range
of possible charm quark momentum distributions.  One sees a similar
behavior with respect to initial production and screening suppression
as was noted in Figure~\ref{jpsicharmdist}.  Thus it is possible to
scan through the initial charm to see the quadratic dependence by
performing an equivalent scan through a corresponding energy range.

\section{Discussion}

The magnitude of enhanced $\J$ yields at RHIC depends on model parameters
which may be subject to some changes as details of the deconfined 
region emerge from data.  However, inclusion of the formation process 
will always lead to an enhancement of $\J$ yields over any calculation
in which dissociation from deconfined gluon collisions plays a role.
Independent of magnitudes, some key characteristics of the
basic physics process will always remain.  Especially significant are 
the quadratic
dependence on initial charm and the increase with centrality. These
features are quite distinct from those obtained by simple extrapolation 
of deconfinement scenarios at SPS energies, and should provide 
model-independent tests.  Since our mechanism is operable only for
the deeply-bound states, the predictions also differ from  
statistical hadronization scenarios.  We also anticipate significant
differences in the momentum spectra, which will be the subject of
future work.

\end{document}